\documentclass[runningheads]{llncs}
\usepackage{graphicx}
\usepackage{amsmath,array,amssymb,amsfonts}
\usepackage{algorithmic}
\usepackage{adjustbox}
\usepackage{color}
\usepackage{hyperref}
\makeatletter
\newcommand{\printfnsymbol}[1]{%
  \textsuperscript{\@fnsymbol{#1}}%
}
\makeatother

\begin{document}
%
\title{Cyclic Generative Adversarial Networks With Congruent Image--Report Generation For Explainable Medical Image Analysis}
\titlerunning{Explainable Medical Image Analysis}
%
\author{Abhineet Pandey\thanks{contributed equally}\inst{1} \and Bhawna Paliwal\printfnsymbol{1}\inst{1} \and Abhinav Dhall\inst{1,2} \and Ramanathan Subramanian\inst{1,3} \and Dwarikanath Mahapatra \inst{4}}
%
\authorrunning{A. Pandey et al.}
%
%
\institute{Indian Institute of Technology (IIT) Ropar, India \and Monash University, Melbourne, Australia \and University of Canberra, Canberra, Australia \and Inception Institute of Artificial Intelligence, Abu Dhabi, United Arab Emirates}

\maketitle              
\begin{abstract}
We present a novel framework for \emph{explainable labeling and interpretation} of medical images. Medical images require specialized professionals for interpretation, and are explained (typically) via elaborate textual reports. Different from prior methods that focus on medical report generation from images or vice-versa, we novelly generate congruent image--report pairs employing a cyclic-Generative Adversarial Network (cycleGAN); thereby, the generated report will adequately explain a medical image, while a report-generated image that effectively characterizes the text visually should (sufficiently) resemble the original. 
The aim of the work is to generate \emph{trustworthy and faithful explanations} for the outputs of a model diagnosing chest x-ray images by pointing a human user to similar cases in support of a diagnostic decision. Apart from enabling transparent medical image labeling and interpretation, we achieve report and image-based labeling comparable to prior methods, including state-of-the-art performance in some cases as evidenced by experiments on the \emph{Indiana Chest X-ray} dataset. 
\keywords{Explainability  \and Medical image analysis \and Multimodal}
\end{abstract}
\section{Introduction}


Medical images present critical information for clinicians and epidemiologists to diagnose and treat a variety of diseases. However, unlike natural images (scenes) which can be easily analyzed and explained by laypersons, medical images are hard to understand and interpret without specialized expertise.

Artificial intelligence (AI) has made rapid advances in the last decade thanks to deep learning. However, the need for accountability and transparency to explain decisions along with high performance, especially in healthcare, has spurred the need for \emph{explainable machine learning}. While natural images can be analyzed and explained by \emph{decomposing} them into semantically-consistent and prototypical visual segments~\cite{Mahapatra_GESTALT_TMI,Mahapatra_Media_SIBNET,KUANAR2022_SP,MahapatraGZSLTMI,LieTMI_2022,Devika_IEEE,MonusacTMI,Mahapatra_Thesis,KuanarVC,MahapatraTMI2021,JuJbhi2020}, as well as \cite{Frontiers2020,Mahapatra_PR2020,ZGe_MTA2019,Behzad_PR2020,Mahapatra_CVIU2019,Mahapatra_CMIG2019,Mahapatra_LME_PR2017,Zilly_CMIG_2016,Mahapatra_SSLAL_CD_CMPB,Mahapatra_SSLAL_Pro_JMI,Mahapatra_LME_CVIU}, multimodal approaches for prototypical explanations are essential for interpreting and explaining medical imagery given the tight connection between image and text in this domain. 

\subsection{Prior Work}
Prior works on medical image interpretation and explainability have either attempted to characterize (chest) x-rays in terms of multiple pathological labels~\cite{LiTMI_2015,MahapatraJDI_Cardiac_FSL,Mahapatra_JSTSP2014,MahapatraTIP_RF2014,MahapatraTBME_Pro2014,MahapatraTMI_CD2013,MahapatraJDICD2013,MahapatraJDIMutCont2013,MahapatraJDIGCSP2013,MahapatraJDIJSGR2013,MahapatraTrack_Book,MahapatraJDISkull2012,MahapatraTIP2012,MahapatraTBME2011,MahapatraEURASIP2010,MahapatraTh2012,MahapatraRegBook} or via automated generation of imaging reports~\cite{mahapatra2022_midl,Mahapatra_CVAMD2021,PandeyiMIMIC2021,SrivastavaFAIR2021,Mahapatra_DART21b,Mahapatra_DART21a,LieMiccai21,TongDART20,Mahapatra_MICCAI20,Behzad_MICCAI20,Mahapatra_CVPR2020,Kuanar_ICIP19,Bozorgtabar_ICCV19} and its variants \cite{Xing_MICCAI19,Mahapatra_ISBI19,MahapatraAL_MICCAI18,Mahapatra_MLMI18,Sedai_OMIA18,Sedai_MLMI18,MahapatraGAN_ISBI18,Sedai_MICCAI17,Mahapatra_MICCAI17,Roy_ISBI17,Roy_DICTA16,Tennakoon_OMIA16,Sedai_OMIA16,Mahapatra_OMIA16,Mahapatra_MLMI16}. The Chexnet framework~\cite{chexnet,Sedai_EMBC16,Mahapatra_EMBC16,Mahapatra_MLMI15_Optic,Mahapatra_MLMI15_Prostate,Mahapatra_OMIA15,MahapatraISBI15_Optic,MahapatraISBI15_JSGR,MahapatraISBI15_CD,KuangAMM14,Mahapatra_ABD2014,Schuffler_ABD2014,Schuffler_ABD2014_2,MahapatraISBI_CD2014,MahapatraMICCAI_CD2013,Schuffler_ABD2013,MahapatraProISBI13,MahapatraRVISBI13,MahapatraWssISBI13,MahapatraCDFssISBI13,MahapatraCDSPIE13} can also be \cite{MahapatraABD12,MahapatraMLMI12,MahapatraSTACOM12,VosEMBC,MahapatraGRSPIE12,MahapatraMiccaiIAHBD11,MahapatraMiccai11,MahapatraMiccai10,MahapatraICIP10,MahapatraICDIP10a,MahapatraICDIP10b,MahapatraMiccai08,MahapatraISBI08,MahapatraICME08,MahapatraICBME08_Retrieve,MahapatraICBME08_Sal,MahapatraSPIE08,MahapatraICIT06} employs a 121-layer convolution network to label chest x-rays.
While the above report-generation works achieve excellent performance, and effectively learn mappings between the image and textual features, they nevertheless do not \emph{verify} if the generated report characterizes the input x-ray. It is this constrained characterization in our suggested work that helps us generate prototypical chest x-ray images serving as explanations. In more recent work saliency maps have been used to select informative xray images \cite{MahapatraTMI2021}


\subsection{Our Approach}
This work differently focuses on the generation of \emph{coherent} image--report pairs, and posits that if the image and report are conjoined counterparts, one should inherently describe the characteristics of the other.
It is the second part of the radiology report generation model i.e. generation of prototypical images from generated reports that serve as explanations for the generated reports. The explainable model proposed can be characterised as a model having post hoc explanations where an explainer outputs the explanations corresponding to the output of the model being explained. The approach to explanations in such an explanation technique as ours is different from methods which propose simpler models such as decision trees that are inherently explainable. Having prototypical images as explanations has been used in case of natural images in \cite{wacv23_Ar,ISR2_Ar,UDA_Ar,IccvGZSl_Ar,ISR_MIDL_Ar,GCN_MIDL_Ar,DevikaAccess_Ar,SouryaISBI_Ar,Covi19_Ar,DARTGZSL_Ar,DARTSyn_Ar,Kuanar_AR2,TMI2021_Ar,Kuanar_AR1,Lie_AR2,Lie_AR,Salad_AR,Stain_AR,DART2020_Ar,CVPR2020_Ar,sZoom_Ar,CVIU_Ar,AMD_OCT,GANReg2_Ar,GANReg1_Ar} can also be \cite{PGAN_Ar,Haze_Ar,Xr_Ar,RegGan_Ar,ISR_Ar,LME_Ar,Misc,Health_p,Pat2,Pat3,Pat4,Pat5,Pat6,Pat7,Pat8,Pat9,Pat10,Pat11,Pat12,Pat13,Pat14,Pat15} and \cite{Pat16,Pat17,Pat18}. None of the approaches explores the paradigm of prototypical image generation as explanations in case of medical images which has been proposed in this work novelly with a multimodal approach.


\subsection{Contributions}
Overall, we make the following research contributions:
\begin{itemize}
\item[1.] We present the first multimodal formulation that enforces the generation of \emph{coherent} and \emph{explanatory} image--report pairs via the cycle-consistency loss employed in cycleGANs~\cite{cyclegan}. 
\item[2.] Different from prior works, we regenerate an x-ray image from the report, and use this image to quantitatively and qualitatively evaluate the report quality. Extensive labeling experiments on textual reports and images generated via the Indiana Chest X-ray dataset reveal the effectiveness of our multi modal explanations approach.
\item[3.] We evaluate the proposed model on two grounds namely: the quality of generated reports and the quality of generated explanations. 
Our method achieves results comparable to prior methods in report generation task, while achieving state-of-the-art performance in certain conditions. The evaluations done for post-hoc explanations show the employability of cycle consistency constraints and multimodal analysis as an explanation technique.
\item[4.] As qualitative evaluation, we present Grad CAM-based \emph{attention maps} conveying where a classification model \emph{focuses} to make a prediction. 
\end{itemize}

\section{Method}~\label{algo}

\subsection{Coherent Image-Report Pairs With CycleGANs}
We aim to model the tight coherence between image and textual features in the chest x-ray images and reports through our multi-modal report generation model. Reports generated should be such that an x-ray image generated with just these generated reports as the input should be similar to ground truth x-ray images; and prototypical x-ray images generated as explanations should be such that a report generated from these images as inputs resembles original report. We hence devise a multimodal, paired GAN architecture explicitly modeling the cycle consistent constraints based on CycleGAN~\cite{cyclegan} with data of type \{\emph{image}, \emph{text}/\emph{labels}\}.

\subsection{CycleGAN}

Given two sets of images corresponding to domains $X$ and $Y$ (for example, two sets of paintings corresponding to different styles), cycleGAN enables learning a mapping $G: X \to Y$ such that the generated image $G(x) = y'$,  where $x \in X$ and $y \in Y$, looks similar to $y$. 

The generated images $y'$ are also mapped back to images $x'$ in domain $X$. Hence, cycleGAN also learns another mapping $F: Y \to X$ where $F(y') =x'$ such that $x’$ is similar to $x$. The structural cycle-consistency assumption is modeled via the cycle consistency loss, which enforces $F(G(x))$ to be similar to $x$, and conversely, $G(F(y))$ to be similar to $y$.
Hence the objective loss to be minimized enforces the following four constraints:
\begin{equation}
G(x) \approx y, F(y) \approx x \hspace{0.25 cm} and \hspace{0.25 cm} F(G(x)) \approx x, G(F(x)) \approx y 
\end{equation}

\begin{figure*}[t]
  \centering
    \leavevmode
      \includegraphics[width = 4.5in, height=2.5in]{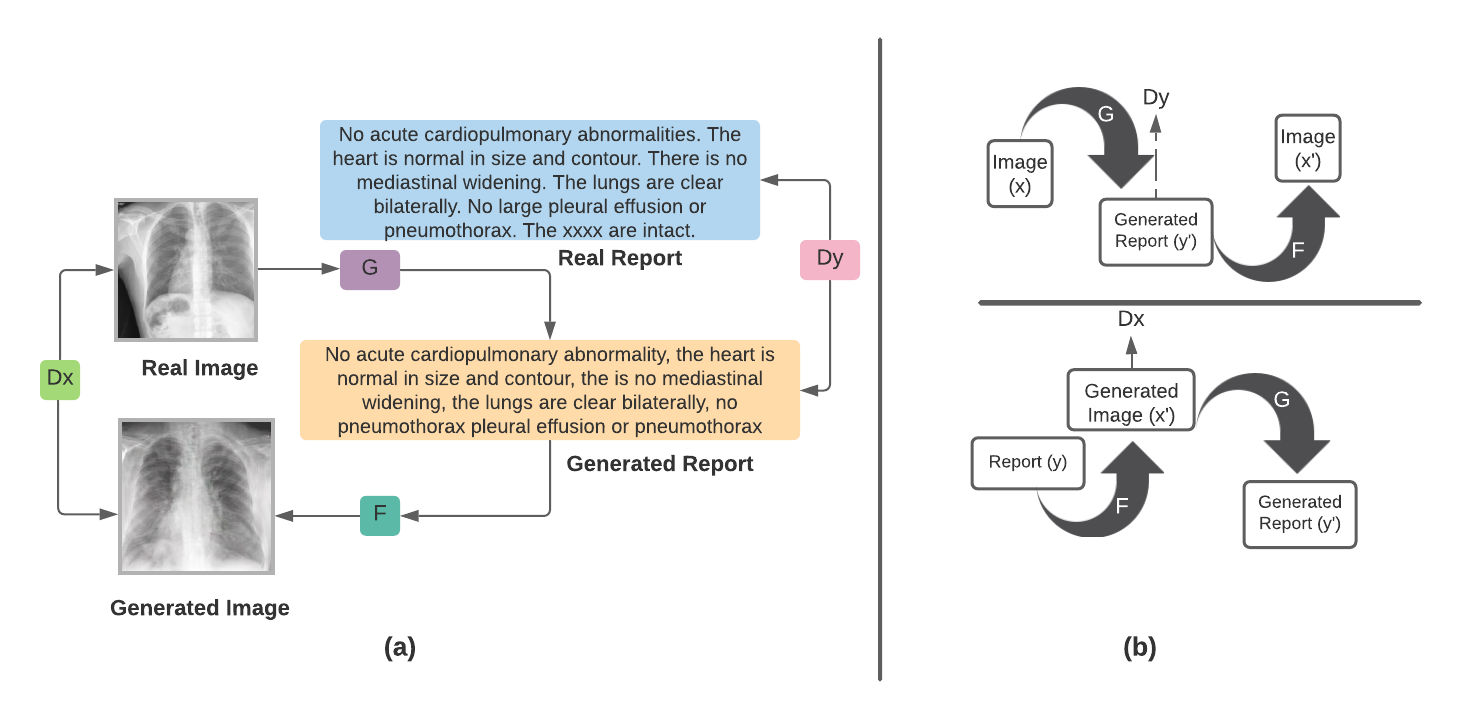}\vspace{-.2cm}
    \vspace{-.4cm}
    \caption{a) Representation of our multimodal cycleGAN framework with exemplar inputs and generated outputs for the image and text modalities. b) Application of the cyclic GAN \cite{cyclegan} framework to generate coherent image--report pairs.}\vspace{-.2cm}
    \label{FigAir}
\end{figure*}

We exploit the setting of Cycle-GAN in a multimodal paradigm i.e. the domains in which we work are text (reports) and image (chest x-ray). As shown in Figure~\ref{FigAir}, our multimodal cyclic GAN architecture comprises (i) two GANs $F$ and $G$ to  respectively generate images from textual descriptions and vice-versa, and (ii) two deep neural networks, termed \emph{Discriminators} $D_X$ and $D_Y$, to respectively compare the generated images and reports against the originals. 
Figure~\ref{FigAir}(a) depicts the mappings $G$ and $F$, while Figure~\ref{FigAir}(b) depicts how cycle-consistency is enforced to generate coherent image-report pairs. 

\subsection{Explanatory image--report pairs}

Our model learns mappings between prototypical image--text decompositions (termed visual or textual words in information retrieval) akin to the \textit{this looks like that} formulation and synthetic image based explanations in . 
Since our setting is multimodal instead of image to image setting in cycle-gans, GAN $G$ (report-to-image generator) in our setting is based on a CNN-plus-LSTM based generative model similar to the architecture proposed . GAN $F$ (image-to-report generation) uses a hierarchical structure composed of two GANs similar to. First, GAN $F_1$ takes the text embedding as input and generates a low-resolution $(64\times 64)$ image. The second GAN $F_2$ utilizes this image and the text embedding to generate a high-resolution $(256 \times 256)$ image. 

\subsection{Dataset}
We used the Indiana University Chest X-Ray Collection (IU X-Ray) for our experiments, as it contains free text reports essential for the report generation task. IU X-Ray is a set of chest x-ray images paired with their corresponding diagnostic reports. The dataset contains 7,470 images, some of which map to the same free text report. 51\% of the images are frontal, while the other 49\% are lateral.


The frontal and lateral images map to individual text reports, at times corresponding to the same report. Consequently, mapping reports to images may confound the generator $F$ regarding which type of image
to generate. To avoid this confusion, we work only with frontal images, thus reducing the dataset to 3793 image-text pairs. Each report consists of the following sections: impression, findings, tags, comparison, and indication. In this work, we treat the contents of impression and findings as the target captions to be generated. We adopt a 80:20 train-test split for all experiments. 

\subsection{Implementation}~\label{implementation}
All images were resized to $244 \times 224$ size. We used $512 \times 512$ images for initial experiments involving the 'Ours-no-cycle' method (see Table~\ref{tab:eval_report}) and observed a better performance with respect to natural language metrics. However, low-resolution x-rays were used for subsequent experiments due to computational constraints. The input and hidden state dimensions for Sentence-LSTM are 1024 and 512 respectively, while both are of 512 length in the case of Word-LSTM. Learning rate used for the visual encoder is 1e-5, while 5e-4 is used for LSTM parts. Embedding dimension used for input to the text-to-image framework is 256, with learning rate set to 2e-4 for both the discriminator and the generator. We used PyTorch-based implementations for all experiments.

Firstly, we individually trained the image-to-text and text-to-image generator modules. In the text-to-image part, we first trained the Stage 1 generator, followed by Stage 2 training on freezing the Stage 1 generator. Note that this individual training of the text-to-image module was done on original reports from the training set. However, when we trained the cycleGAN architecture, the text-to-image part took in the generated text as input. While directly training both the modules together, oscillations in loss values were observed.

\begin{table}[t]
\caption{Natural Language Metrics for Generated Reports}\label{tab:eval_report}
\begin{center}
\begin{tabular}{ |c|m{2 cm}|m{2 cm}|m{2 cm}|m{2 cm}| } 
\hline
Methods: & Ours-Cycle$^*$ & Ours-no-cycle & R2Gen  & Multiview \\
\hline
BLEU-1 & 0.486 & 0.520 & 0.470 & \textbf{0.529}\\ 
BLEU-2 & 0.360 & \textbf{0.388} & 0.304 & 0.372 \\ 
BLEU-3 & 0.285 & 0.302 & 0.219 & \textbf{0.315} \\
BLEU-4 & 0.222 & 0.251 & 0.165 & \textbf{0.255} \\
ROUGE & 0.440 & \textbf{0.463} & 0.371 & 0.453 \\
\hline
\end{tabular}
\end{center}
\footnotesize{$^*$ Reduction in training data (only frontal image-report pairs used)}\\ 
\vspace{-5mm}
\end{table}

\section{Evaluation}

\subsection{Evaluation of Generated Reports}

We first evaluate the quality of the generated reports via the BLEU and ROUGE metrics we compare our performance against other  in Table~\ref{tab:eval_report}. Our methods with and without cycle-consistency loss are referred to as \emph{Ours-cycle} and \emph{Ours-no-cycle}. Since only frontal images were used for training \textit{Ours-cycle} (see Section~\ref{implementation}), the training set is reduced to 3793 image--report pairs. We get comparable performance with the multi-view network~\cite{multiview} based on NLG metrics. There is a small drop in these metrics with the addition of the cycle component, mainly due to the reduction in training data (as  the number of image-report repairs is approximately halved).

\subsection{Evaluation of Explanations}
To evaluate the explanations, we first assess if the generated images truly resemble real input images because the quality of the generated images is also a representative of the quality of the model generated reports as discussed in earlier sections. Secondly, we consider the aspects of trust and faithfulness of our explanation technique based on ideas in  for post-hoc explanations. 
\subsubsection{3.2.1 Evaluating Similarity of Generated Images and Real X-ray Images}
We quantitatively assess the images using CheXNet \cite{chexnet}  (state-of-the-art performance on multi-label classification for chest x-ray images). We use CheXNet on $\langle$input image--generated image$\rangle$ pairs for checking the amount of disparity present between the \emph{true} and \emph{generated} images. We achieve a KL-Divergence of 0.101. We also introduce a 'top-k' metric to identify if the same set of diseases are identified from the \emph{input} and \emph{generated} images. The metric averages the number of top predicted diseases which are \emph{common} to both input and the generated images.\newline
\begin{center}

        $top-\textit{k} = \frac{\sum_{All\,pairs}\mid(top-\textit{k}\;labels(input\,image)) {\cap} (top-\textit{k} \;labels(generated\,image))\mid}{Number\,of\,pairs}$

\end{center}
We compare the output labels of CheXNet on both real and generated image using the top-$k$, Precision@$k$ and Recall@$k$ metrics. From Table \ref{table:nonlin}, on average 1.84 predicted  disease labels are common between the input and generated images, considering only the top-two ranked disease labels. In Table \ref{table:nonlin}, we have also shown a comparison against images generated from our text-to-image (report-to-x-ray-image) model on the reports generated by the recently proposed transformer-based R2gen algorithm. Our representative generated images perform better on  the top-x, precision and recall metrics, quantitatively showing that the reports generated by our cycleGAN model better describe the input chest x-ray image.

\begin{table}[t]
\caption{Metrics for generated images by using CheXNet for multi-label classification.} 
\centering 
\begin{adjustbox}{width=\textwidth}
\begin{tabular}{|c | m{2 cm} | m{2 cm} | m{2 cm} | m{2 cm} | m{2 cm} | m{2 cm} | m{2 cm} |} 
\hline
k & Top-k (Ours) & Top-k (R2Gen) \cite{transformer} & Precision@k (Ours) & Precision@k (R2Gen) \cite{transformer} & Recall@k (Ours) & Recall@k (R2Gen) \cite{transformer}\\ [0.5ex] 
\hline 
2 & \textbf{1.84} & 0.64 & \textbf{0.92} & 0.32 & \textbf{0.13} & 0.05\\
5 & \textbf{3.01} & 2.55 & \textbf{0.60} & 0.51 & \textbf{0.21} & 0.18\\
8 & \textbf{6.45} & 5.82 & \textbf{0.81} &  0.73 & \textbf{0.46} & 0.42\\ [1ex]
\hline
\end{tabular}
\end{adjustbox}
\label{table:nonlin} 
\end{table}




\subsubsection{3.2.2 Evaluating Trustability of the Explanations}
We build upon the idea of trust in an explanation technique suggested in  for post-hoc explanations. An explanation method can be considered trustworthy if the generated explanations are able to characterize the kind of inputs on which the model performs well or closer to the ground truth. We evaluate our explanations on this aspect of trustability by testing if the explanations or prototypical x-ray images generated are the images on which reports generated are very close to ground truth reports. We evaluate the similarity of the two reports (ground truth reports and reports generated from prototypical images) by comparing the labels output by a naive Bayes classifier on the input reports. The results for accuracy metric for each of the 14 labels is summarised in the Table 3. We can clearly infer that the x-ray images generated as explanations have been able to understand the model's behaviour and hence the good accuracy (around 0.9 for most of the labels). 

\begin{table}[t]
\caption{Accuracy Metric for the Reports Generated from Prototypical (Generated) Images} 
\centering 
\begin{adjustbox}{width=\textwidth}
\begin{tabular}{|c | m{0.8 cm} | m{0.8 cm} | m{0.8 cm} | m{0.9 cm} | m{0.8 cm} | m{0.95 cm} | m{0.8 cm} | m{0.8 cm} | m{0.8 cm} | m{0.8 cm} | m{0.8 cm} | m{0.8 cm} | m{0.8 cm} | m{0.8 cm} | m{0.8 cm} |} 
\hline
\rule{0pt}{29pt}Label & No Finding & Cardi omed iast inum & Cardi omeg aly & Lung Lesion & Lung Opac ity & Edema & Cons olida  tion & Pneu mon ia & Atele ctasis & Pneu moth orax & Pleur al(E) & Pleur al(O) & Fract ure & Supp ort Devic es\\ [0.5ex] 
\hline 
\rule{0pt}{15pt}Accuracy & 0.78 & 0.92 & 0.84 & 0.96 & 0.82 & 0.97 & 0.96 & 0.97 & 0.94 & 0.98 & 0.95 & 0.99 & 0.96 & 0.94\\ [1ex]
\hline
\end{tabular}
\end{adjustbox}
\label{table:nonlin2} 
\end{table}
\subsubsection{3.2.3 Evaluating Faithfulness of Explanations}
Another aspect which has been explored in some of the explanation works is faithfulness of the technique i.e. whether the explanation technique is reliable. Reliability is understood in the sense that it is reflecting the underlying associations of the model rather than any other correlation such as just testing the presence of edges in object detection tasks~. We test the faithfulness of the explanations generated by randomising the weights of the report generation model and then evaluating the quality of prototypical images to check if the explanation technique can be called faithful to the model parameters. The metric values for Top-2, Precision@2 and Recall@2 for generated images in this case are 0.90, 0.45 and 0.06 respectively significantly less than corresponding metrics in Table 2. As evident, the prototypical images generated as explanations from randomised weights model are unable to characterize the original input images because the model they are explaining doesn't contain the underlying information it had previously learnt for characterizing given chest x-ray images. 


\subsubsection{3.2.4 Qualitative Assessment of Generated Images using Grad-CAM}
\begin{figure*}[ht]
  \centering
    \leavevmode
      \includegraphics[width = \linewidth]{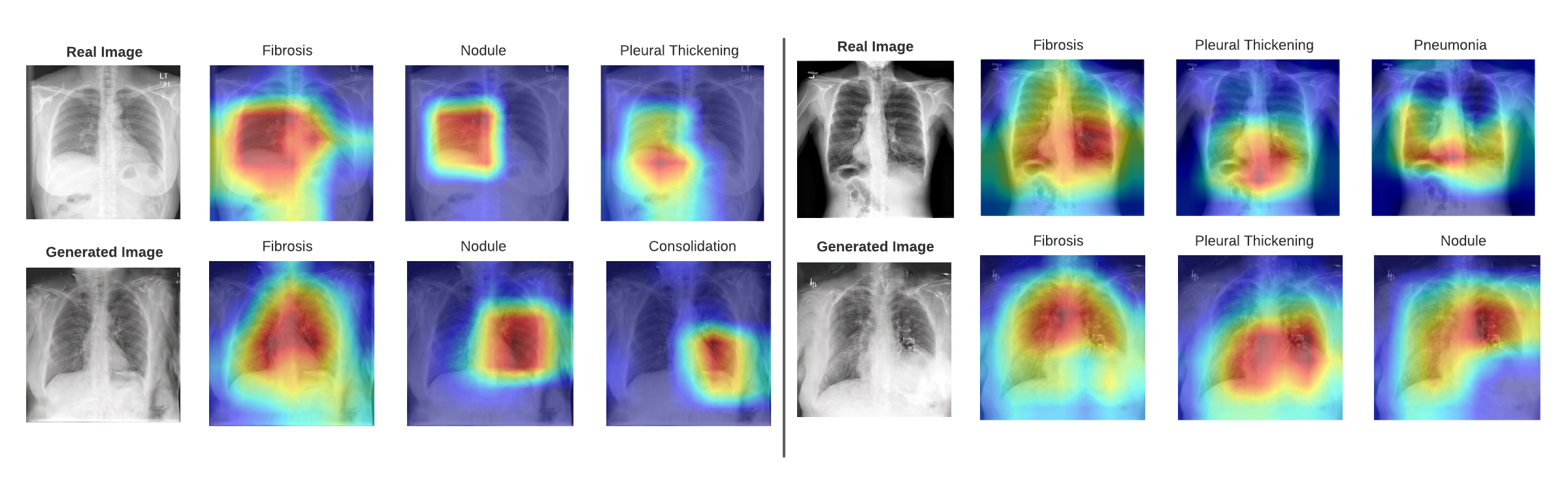}
    \caption{Grad CAM saliency maps for top 3 predicted labels by CheXNet for real (top row)  and generated (bottom row) image pairs; Sample pair 1 (left) and Sample pair 2 (right)}
    \label{FigAir3}

\end{figure*}

We used GradCAM  for highlighting the salient image regions focused upon by the CheXNet \cite{chexnet} model for label prediction from the real and generated image pairs. Two examples are shown in Fig.~\ref{FigAir3}. In the left sample pair, real image shows fibrosis as the most probable disease label, as does the generated image. As observable, the highlighted region showing the presence of a nodule is the same in both x-ray images except for the flip from the left and right lung. This shows that the report generation model was able to capture these abnormalities with great detail, as the report-generated image also captures these details visually. Similarly, two of the top-three labels are the same in both real and generated images as predicted by CheXNet in sample pair 2. 


\section{Conclusion}
A cycleGAN-based framework for explainable medical report generation and synthesis of coherent image-report pairs is proposed in this work. 
Our generated images visually characterize the text reports, and resemble the input image with respect to pathological characteristics. We have performed extensive experiments and evaluation on the generated images and reports, which show that our report-generation quality is comparable to the state-of-the-art in terms of natural language generation metrics; also the generated images depict the disease attributes both via attention maps and other quantitative measures (precision analysis, trust, and faithfulness) showing the usefulness of a cycle-constrained characterization of chest x-ray images in an explainable medical image analysis task.

\bibliographystyle{splncs04}
\bibliography{ms}

\end{document}